\documentclass[final,3p,times]{elsarticle}
\usepackage{hyperref}
\usepackage{xcolor}
\usepackage{amssymb}
\usepackage{rotating}
\usepackage{booktabs}
\usepackage{wrapfig}
\usepackage{adjustbox}
\usepackage{algorithmic}
\usepackage{algorithm2e}
\RestyleAlgo{ruled}
\usepackage{amsmath}
\usepackage{svg}
\newcommand*{\Scale}[2][4]{\scalebox{#1}{$#2$}}%
\usepackage{multirow}
\newcommand{\orcid}[1]{\href{https://orcid.org/#1}{\includegraphics[width=10pt]{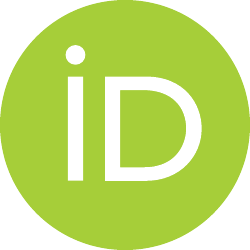}}}

\journal{Applied Soft Computing}

\begin{document}

\begin{frontmatter}

\title{An Adaptive and Altruistic PSO-based Deep Feature Selection Method for Pneumonia Detection from Chest X-Rays}

\author[inst1]{Rishav Pramanik \orcid{0000-0003-0144-8539}}

\affiliation[inst1]{organization={Department of Computer Science and Engineering},
            addressline={Jadavpur University}, 
            city={Kolkata},
            postcode={700032},
            country={India}}

\author[inst2]{Sourodip Sarkar \orcid{0000-0002-7050-5835}}
\affiliation[inst2]{organization={Department of Electronics and Communication Engineering},
            addressline={Heritage Institute of Technology}, 
            city={Kolkata},
            postcode={700107}, 
            country={India}}

\author[inst1]{Ram Sarkar \orcid{0000-0001-8813-4086}}
\nonumnote{Corresponding\;author:\;Rishav\;Pramanik,\\E-mail:\;rishavpramanik@gmail.com}

\begin{abstract}
Pneumonia is one of the major reasons for child mortality especially in {income-deprived} regions of the world. Although it can be detected and treated with very less sophisticated instruments and medication, Pneumonia detection still remains a major concern in developing countries. Computer-aided based diagnosis (CAD) systems can be used in such countries due to their lower operating costs than professional medical experts. In this paper, we propose a CAD system for Pneumonia detection from Chest X-rays, using the concepts of deep learning and a meta-heuristic algorithm. We first extract deep features from the pre-trained {ResNet50,} fine-tuned on a target Pneumonia dataset. Then, we propose a {feature selection} technique based on particle swarm optimization (PSO), which is modified using a memory-based adaptation parameter, and enriched by incorporating an altruistic {behavior} into the agents. We name our {feature selection} method as adaptive and altruistic PSO (AAPSO). The proposed method successfully eliminates non-informative features obtained from the ResNet50 model, thereby improving the Pneumonia detection ability of the overall framework. Extensive experimentation and thorough analysis on a publicly available Pneumonia dataset establish the superiority of the proposed method over several other frameworks used for Pneumonia detection. {Apart from Pneumonia detection, AAPSO is further evaluated on some standard UCI datasets, gene expression datasets for cancer prediction and a COVID-19 prediction dataset. The overall results are satisfactory, thereby confirming the usefulness of AAPSO in dealing with varied real-life problems. The supporting source codes of this work can be found at~\href{https://github.com/rishavpramanik/AAPSO}{https://github.com/rishavpramanik/AAPSO}.}
\end{abstract}


\end{frontmatter}

\section{Introduction}
\label{sec:introduction}
\begin{figure*}
    \centering
    \includegraphics[width=\linewidth,height=2.8cm,keepaspectratio]{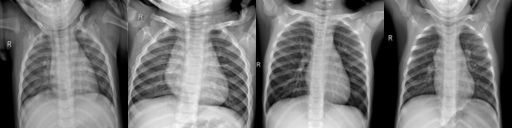}
    \includegraphics[width=\linewidth,height=2.8cm,keepaspectratio]{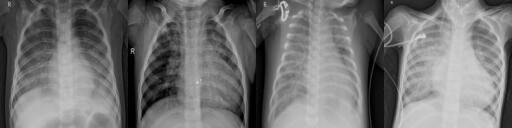}
    \caption{CXRs of Pneumonia and normal cases. Images are taken from the Kermany et al.~\cite{kermany2018identifying} dataset. The top row consists of images belonging to the normal category, whereas images in the bottom row belong to confirmed Pneumonia cases.}
    \label{fig:dataset}
\end{figure*}

Pneumonia is {a} very common {disease,} especially among children. {However,} Pneumonia can be treated with {low-cost} medication. Despite having some affordable treatment procedures, it is very unfortunate that majorly in sub-Saharan Africa and South Asia where poverty still exists to a large extent, it continues to be a reason to cause thousands of deaths every year.\footnote{\href{https://www.who.int/news-room/fact-sheets/detail/Pneumonia}{https://www.who.int/news-room/fact-sheets/detail/Pneumonia}}. A possible reason for this is the lack of infrastructural {facilities,} such as proper testing labs. Besides, pollution and the lack of sense of hygiene due to lower levels of education make things worse. 
A Chest X-Ray (CXR) impression of the lung area is one of the most effective ways to detect Pneumonia~\cite{makhnevich2019clinical}. A beam of radiation is passed through the human body, and the image is collected on a special film. {Thus,} the entire process of examination becomes completely painless and quicker than methods like Computed Tomography (CT) scans. Sample CXRs of Pneumonia {affected} and normal cases are {illustrated} in Fig.~\ref{fig:dataset}. {Typically,} there is a good amount of intra-class and inter-class variations for CXRs {that make} the computer-based detection task much more challenging. This is {because} the physical structure of every human being is different. The range of Pneumonia patches depends on the severity of infection. Further, the developed Pneumonia patches may have varied {shapes} and can be located in multiple areas~\cite{ho2015usefulness}. This makes the task of detection even more difficult.\\
With the advancement of technologies, Computer-Aided Detection (CAD) systems are nowadays getting more popular because it requires much less investment for medical laboratory setup, thereby making the medical facilities affordable for all. {Typically,} such systems consider an image as the input of the suspected organ to make the prediction. Computer scientists generally extract a set of features from the inputs by some means and {try to} identify the presence of a disease using some machine learning algorithms. Deep learning-based methods now do the same, but they do not need any feature engineering for the detection or classification tasks. Researchers usually perform deep feature extraction with the help of convolutional neural networks (CNNs) and classify using multi-layered neural networks. To train such networks, they rely on an objective function, which is used to optimize the numerous internal-weight values of the deep model. Hence, this is known as \textit{“hilly landscape of multiple weight values”}~\cite{lecun2015deep}. Deep learning-based methods have shown good generalization traits over {various problem domains}, which prompts researchers around the globe to work tirelessly and come up with more efficient and effective models than earlier. However, this robust nature comes at the cost of high computational resources {and, in general it requires a huge amount of data to train the model efficiently. The latter requirement} sometimes cannot be fulfilled, especially in the biomedical field.\\
{Nowadays,} researchers very often use the concept of transfer {learning,} which alleviates the need for such a huge amount of data for proper training of the models. For this they train the network using a larger dataset (such as ImageNet), then {they transfer weights and using which they} train the network on a smaller dataset (i.e., the target dataset). However, researchers generally overlook one important aspect when they apply deep learning models, which is the existence of redundant or non-informative features, the presence of which may hamper the overall performance of the network. For that reason, recently we see a good influx of {pruning-based} deep networks which essentially aim to eliminate irrelevant parts of the deep network and re-train the network. On a similar note, feature selection-based algorithms aim for the same, i.e., such algorithms focus on selecting the best subset of features from a given set of features. {Researchers, often devise such methods using popular meta-heuristic-based} optimization algorithms such as Genetic Algorithm (GA), Particle Swarm Optimization (PSO), and Ant Colony Optimization (ACO) to name a few.\\
In the present work, we propose an adaptive and altruistic PSO (AAPSO) for Pneumonia detection from CXR images. We first use a CNN architecture (pre-trained on the ImageNet dataset), which {we fine-tune} on the target dataset. The features from the layer preceding the final classification layer are extracted and fed to the proposed AAPSO algorithm to obtain the relevant set of features only, and finally, classify {the CXR images} using the $k-$ Nearest Neighbors ($k-$NN) classifier. The proposed deep {feature-selection} method outperforms several state-of-the-art approaches. In a nutshell, {we list} the highlights of our work as:
\begin{itemize}
   \item {We propose a deep feature selection-based method for Pneumonia detection from CXR images.}
   \item {Original PSO is improvised with a dynamic adaption parameter which we propose on the assumption: \textit{``Relative divergence of a solution is directly dependent on the search time remaining."}}
   \item {Adaptive PSO is further enriched with altruistic qualities for the motive to select the most relevant subset of features from a pool of features extracted by the CNN model.}
   \item {The proposed deep feature selection framework is assessed on a publicly accessible Pneumonia CXR dataset using 5-fold cross-validation scheme.}
   \item {The proposed framework is also tested on several real-life datasets like gene expression-based cancer prediction and COVID-19 prediction to ensure the robustness of the same.}
\end{itemize}
The rest of the paper is organized in the following manner: Section \ref{sec:related} consists of {past methods proposed} in the domain of Pneumonia detection from CXRs and also reviews the different variants of PSO found in the literature for feature selection. Section \ref{sec:proposed} gives a detailed description of the proposed method, and section \ref{sec:results} reports the experiments and the corresponding analysis. To check the usefulness of AAPSO on other real-life medical datasets, in section~\ref{sec:real} we extensively evaluate AAPSO on standard UCI datasets, {microarray-based} gene expression datasets and a COVID-19 prediction dataset. Finally, we conclude our paper in section \ref{sec:conclusion}.
\section{Related Work}
\label{sec:related}
\subsection{Pneumonia Detection From CXRs}
Some recently proposed methods {for} Pneumonia detection using CXRs are discussed below:\\
Liang and Zheng~\cite{liang2020transfer} proposed a deep residual-based model with dilated convolutions {having} 49 layers for Pneumonia detection. {Besides, they added some noise} to deal with overfitting. While this inclusion was able to handle the overfitting problem, on the other hand, the experimental results show that the method is less precise for Pneumonia detection which may not be useful in practical scenarios. This problem was possibly due to the reason that the authors opted to adapt the transfer learning procedure by training the network on a large-scale CXR dataset~\cite{wang2017chestx}. The dataset used for pre-training itself is imbalanced, therefore there might have been a problem to learn the feature representation in the pre-training process itself. Recently, Zhang et al.~\cite{zhang2021viral} proposed a one-class detection technique. {The authors used a deep model for feature extraction, and then} proposed an anomaly-detection module and a confidence-prediction module, and to fit the anomaly scores the authors used Gaussian distribution. Although the method is of great significance, the authors directly used a pre-trained feature extractor. A {fine-tuned} network might have given a better feature {representation,} which would have been {more useful} for final classification.\\
The work by Chattopadhyay et al.~\cite{chattopadhyay2021Pneumonia} proposed a deep feature-selection technique with a Sine-Cosine Algorithm aided by a local search method. 
Unsupervised learning approaches have also been proposed in the past. For example, Tang et al.~\cite{tang2019tuna} aimed to evaluate the classification-based generalizability of Generative Adversarial Network (GAN) based methods. The authors proposed to use CycleGAN for this task. The authors used the feature maps from the inner layers to calculate the reconstruction loss for unsupervised learning.\\
Ensemble-based methods such as the one by Kundu et al.~\cite{kundu2021Pneumonia} used three deep CNN models as base learners, and outcomes of those learners were aggregated based on {the} weighted sum rule. The {authors assigned weights based on the} entropy of various performance metrics. One of the main problems with this strategy is that deep learners often give very high probabilistic values even for misclassification scenarios. {This behavior is observed due to} irrelevant features learned within its inner layers. The authors assigned the classifier a weight, used for classification, which might not be practical considering the previous discussion. In the article by Dey et al.~\cite{dey2021customized}, the authors proposed an ensemble scheme using principal component analysis (PCA) and a serial fusion. The method was designed to ensemble deep features and handcrafted features such as Complex Wavelet Transform (CWT), Discrete Wavelet Transform (DWT) and Gray-Level Co-Occurrence Matrix (GLCM). The {ensembled features were classified }using some popular machine learning-based classifiers. {A high-dimensional} feature vector was used for {the} final classification. This might have been a concern in performance since no feature selection technique was applied to check the existence of correlated features.
\subsection{Feature Selection Using PSO}
Since its inception in 1995~\cite{kennedy1995particle}, PSO has been used successfully to handle many real-life continuous optimization problems~\cite{bi2020energy,yuan2016ttsa}. Additionally, for a discrete search space-based optimization problem, like the knapsack problem or the feature selection problem, PSO has been used by many researchers. {Below, we review some} variants of PSO proposed in recent {years.}\\
Tran et al.~\cite{tran2018variable} proposed a Variable-Length Particle Swarm Optimization (VLPSO), which divided the population into several partitions. Each partition had a maximum length criterion, which {equals} the number of features selected from a subset of ranked features. This ranking was done based on symmetrical uncertainty – a filter method. The proposed strategy could very well reduce the computational memory and time required. However, over the course of iterations, the feature ranking was not updated as it is a computationally expensive task. But it may hamper the overall feature selection process. In addition, the particles were updated independently in the different dimensions without considering the interactions among themselves. In {another} work by Tran et al.~\cite{tran2017new}, they proposed potential PSO (PPSO), a {feature-representation} mechanism, which could perform feature discretion and selection. This strategy reduced the search space complexity. A new fitness function was {also} proposed to evaluate the solutions. {Results} achieved on standard datasets were encouraging, but in the initial stages the method uses a fixed size feature vector for random {initialization,} this may pose a challenge to explore the search space. Also, this method requires a pre-defined list of cut-points, which might pose an additional challenge.\\
Ansari et al.~\cite{ansari2019hybrid} proposed a bi-stage feature selection method where the first stage consists of two filter methods, and the second stage consists of two {wrappers for feature-selection}. This approach resulted in using a lot of computational resources and time to optimize with four feature selection algorithms. In the work by Ghosh et al.~\cite{ghosh2020binary}, the authors used GA and PSO, then applied the average weighted combination method (AWCM) followed by a local search method namely sequential one-point flipping (SOPF). This approach was useful and resulted in a good number of feature reductions. The proposed framework requires determining a feature importance score based on {accuracy values} obtained by GA and PSO, and using the mean of this importance {score,} features were eliminated. Therefore, the dependency of one feature on another was completely ignored. A work proposed by Guha et al.~\cite{guha2021hybrid} proposed to hybridize PSO and Gravitational Search Algorithm (GSA) for handwritten script classification. The proposed method introduced a concept of memory into the memory-less GSA while updating the velocity of the agents. But the driving force {for} velocity i.e., the acceleration, which is a measure used in GSA to determine the extent of exploration and exploitation, remains memory free in their proposed algorithm. Hence, the algorithm may have some issues balancing the exploration and exploitation {efficiently.}\\
\section{Methods and Materials}
\label{sec:proposed}
\begin{figure*}[ht!]
    \centering
    \includegraphics[width=\linewidth,keepaspectratio]{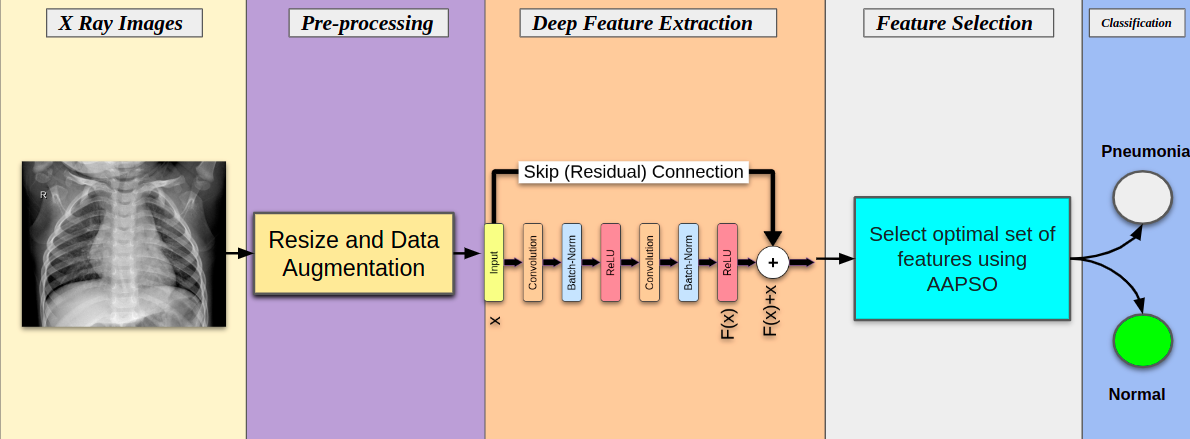}
    \caption{The overall pipeline of the proposed approach used for Pneumonia detection from CXRs. First, the images are resized and augmented using standard techniques. Then these are fed to the pre-trained ResNet50 model for deep feature extraction. After that the proposed AAPSO selects the most informative features. Finally, the classification is performed using the $k-$NN classifier.}
    \label{fig:pipe}
\end{figure*}

In this {section, we discuss} the relevant details corresponding to the proposed deep feature selection framework for Pneumonia detection from CXRs. {We present} the overall pipeline of the proposed approach in Fig.~\ref{fig:pipe}.
At first, the input images are resized to (224,224) pixels and augmented using standard online data augmentation techniques such as horizontal and vertical flips, rotation, scaling, skew and translation. This step is essential to make the model capable of dealing with input variants~\cite{pramanik2022fuzzy}. However, for the test {set, the data augmentation process} is avoided to ensure the model is not evaluated on synthetic data. These samples are then fed to the ResNet50 model (pre-trained on the ImageNet dataset) and are used to train the deep CNN, for classification. We use fully-connected neural networks, essentially with one hidden layer of the dimension of 512. The last layer is the probability score generator, hence we extract deep features from the {second-last} layer (512 dimensions). These features are fed to the proposed {AAPSO-based} feature selection algorithm to remove redundant and non-informative features, and thus a reduced feature set is obtained. This reduced feature set is fed to the $k-$NN classifier for the {purpose of classification}.
\subsection{Dataset Description}
We have used a publicly available CXR Pneumonia dataset by Kermany et al.~\cite{kermany2018identifying}. We should be cognizant of the fact that the dataset suffers from the class-imbalance problem. 
This dataset is publicly-hosted on the Kaggle platform\footnote{\href{https://www.kaggle.com/datasets/paultimothymooney/chest-xray-pneumonia}{https://www.kaggle.com/datasets/paultimothymooney/chest-xray-pneumonia}} for easy use. We present the distribution of the images in Table~\ref{tab:data}. 
\begin{table}[ht!]
    \centering
        \caption{Distribution of images in the Pneumonia dataset used for experimentation.}
    \begin{tabular}{c|c|c}
        \toprule
        Class & Setting & Samples \\
        \midrule
        \multirow{2}{*}{Normal} & Train & 1267\\
        & Test & 316\\
        \midrule
       \multirow{2}{*}{Pneumonia} & Train & 3419\\
        & Test & 856\\
        \bottomrule
    \end{tabular}
    \label{tab:data}
\end{table}

\subsection{Deep Residual Network}
\begin{figure}[ht!]
    \centering
    \includegraphics[width=\linewidth,height=7cm,keepaspectratio]{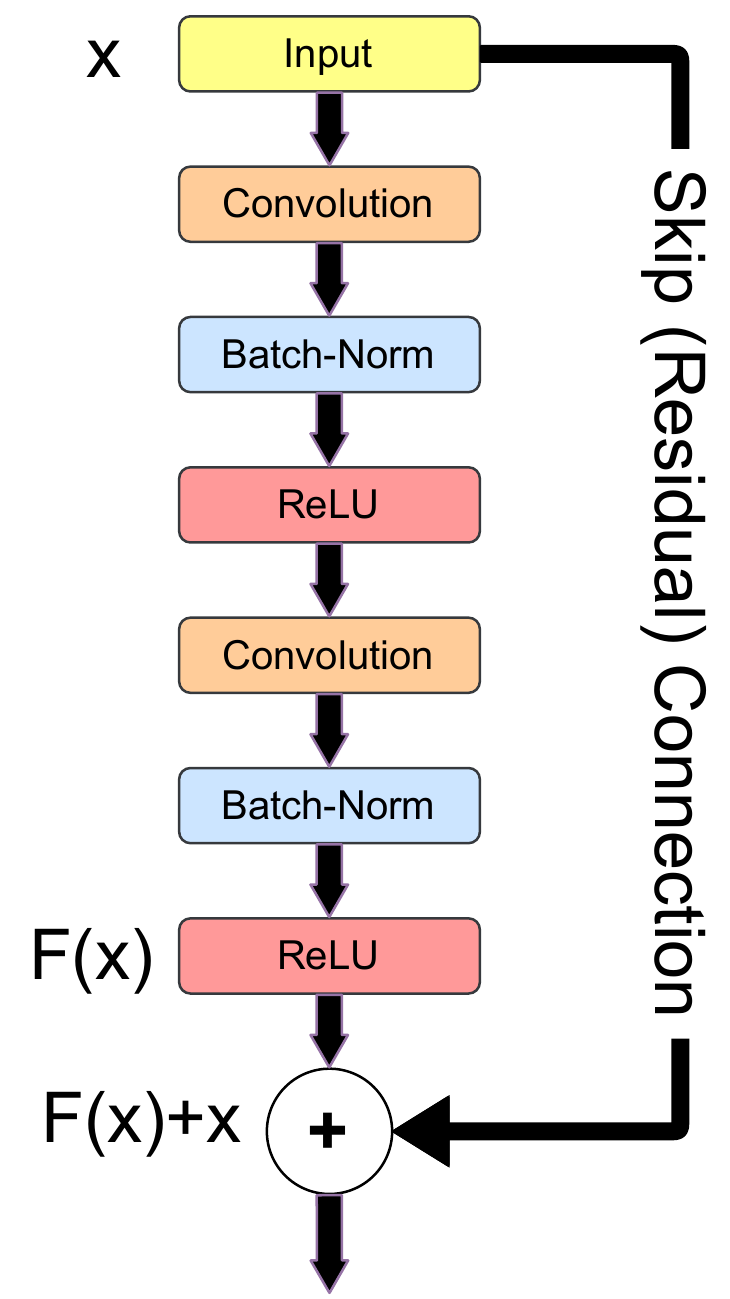}
    \caption{A typical residual block used in ResNet-based deep CNNs. Modified from \cite{he2016deep}}
    \label{fig:resnet}
\end{figure}

Deep Residual Network (ResNet), one of the widely used deep learning {models,} was introduced by He et al.~\cite{he2016deep} in 2016. {Typically,} a ResNet architecture {comprises} several residual blocks. The very first thing we can notice in Fig.\ref{fig:resnet} is that there is a direct connection that skips some layers of the model. The {residual-connections} help to counter the vanishing gradient {problem,} which means {for a} very deep CNN architecture, the gradients, or the derivatives tend to be zero and thus are not properly optimized during the training process. For a deep architecture, a signal is generally processed {following} Equation~\ref{eqn:resnet1} where $w$ is referred to as the weight parameter and $b$ is called the {bias,} $w$ and $b$ are trainable. {With} residual learning, the input signal is modified following Equation~\ref{eqn:resnet2}.
\begin{gather}
    \label{eqn:resnet1}
    F(x)=ReLU(wx + b)\\
    \label{eqn:resnet2}
    H(x)=F(x)+x
\end{gather}
In addition, the residual connections regularize the model. Suppose a block gives a weak representation of features. For such a scenario, residual connections would eventually retain the original representation, thereby making the model much {less} prone to over-fitting. In {the present} work, we use {ResNet50,} which consists of 4 {residual} convolutional blocks (Fig.~\ref{fig:resnet}).
\subsection{Particle Swarm Optimization}
PSO, originally introduced by Kennedy and Eberhart~\cite{kennedy1995particle}, is a {population-based} optimization algorithm. The success of PSO over the years lies in the fact that it is simple, {has fewer} controlling {parameters,} and is computationally inexpensive in terms of memory use. The mathematical formulation of PSO is given in Equations \ref{eqn:PSO1},\ref{eqn:PSO2}.
\begin{gather}
    \label{eqn:PSO1}
    \Scale[0.75]{
    v_{ij}(t+1)=v_{ij}(t)+r1*(Pbest_{ij}-x_{ij}(t))+r2*(Gbest_{j}-x_{ij}(t))}\\
    \label{eqn:PSO2}
    x_{ij}(t+1)=x_{ij}(t) + v_{ij}(t+1)
\end{gather}
In Equation \ref{eqn:PSO1}, $r1$ and $r2$ are two random numbers in the range of $(0,1)$, and $v_{ij}$ and $x_{ij}$ refer to the velocity and the position respectively for the $i^{th}$ particle in $j^{th}$ dimension. $Pbest$ is the personal best solution of the given agent. $Gbest$ is the global best solution derived from the global best agent.
\subsection{Adaptive PSO}
\subsubsection{Motivations Behind Adaptive PSO}
The basic PSO algorithm sometimes suffers from a major limitation which {is,} if the population hovers around a particular {suboptimal} solution, the chances of getting stuck in such a particular region of the search space become much higher. Also, the algorithm does not consider the extent of exploration and exploitation that is {desired} to reach the global optima. To this end, we hypothesize: \textit{“Relative divergence of a solution is directly dependent on the search time remaining.”} This is true when we consider the maximum number of iterations as the stopping criterion. {The justification behind the hypothesis is that in initial iterations, the search process should go through more diverse regions, thereby ensuring proper exploration. With progression in time, the particles try to converge to an optimal solution, thereby ensuring exploitation in the later iterations. To address this issue, we incorporate a memory-based adaptive dependence parameter into the basic PSO algorithm.}\\
\subsubsection{Derivation of The Adaptation Parameter}
{Equation~\ref{eqn:PSO-1} gives the mathematical foundation of the hypothesis stated above. In this equation, the term $\frac{dS}{S}$ quantifies our assumption of relative divergence from a solution. {As we consider,} $dS$ refers to the change in a solution, whereas $S$ is the present {solution. Hence} by relative {divergence,} we measure the change in the solution w.r.t. the present solution. The change in time is represented by $dt$.}
\begin{gather}
    \label{eqn:PSO-1}
    \frac{dS}{S} \propto dt\\
    \label{eqn:PSO-2}
    \frac{dS}{S}=c\cdot dt\\
    \label{eqn:PSO-3}
    \int_{S_i}^{S_f} \frac{dS}{S}=c\int_{t_i}^{t_f} dt\\
    \label{eqn:PSO-4}
    \log_e {\frac{S_f}{S_i}} = c \cdot (t_f-t_i)\\
    \label{eqn:PSO-5}
    S_f = S_i\cdot e^{c(t_f-t_i)} \;\;\; \Scale[0.75]{(\Delta S=S_i-S_f)} \\
    \label{eqn:PSO-6}
    \frac{\Delta S}{S_i}= 1-e^{c(t_f-t_i)}\\
    \label{eqn:newPSO}{
    v_{ij}(t+1)=w*v_{ij}(t)+r1*(Pbest_{ij}-x_{ij}(t))+r2*(Gbest_{j}-x_{ij}(t))}
\end{gather}
{In Equation~\ref{eqn:PSO-2}, we introduce an equality term amongst the assumed {relations} using a proportionality constant of $c$. For the experimental purpose, the value of $c$ is set to $1$. Further, following the basic rules of integration, we integrate within {limits. We set the limits} keeping in mind the start and end of the algorithm w.r.t. time, {which is here} represents the number of iterations. $S_i$ refers to the initial solution, whereas $S_f$ refers to the final solution. In Equation~\ref{eqn:PSO-4}, we apply the relevant limits. After some rearrangements as shown in Equation~\ref{eqn:PSO-5}, finally in Equation ~\ref{eqn:newPSO} we get the value of $w$ as derived in Equation~\ref{eqn:PSO-6}.}
\subsection{Transfer Function and Fitness Value}
\subsubsection{Transfer Function}
\begin{figure}
    \centering
    \includegraphics[width=\linewidth,height=5cm,keepaspectratio]{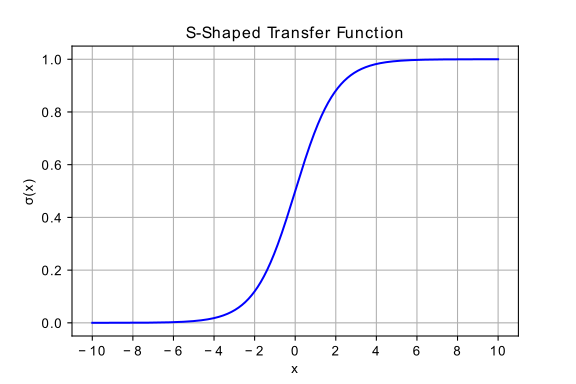}
    \caption{A graphical representation of S-shaped transfer function. This function is commonly known as the sigmoid function}
    \label{fig:transfer}
\end{figure}

As discussed earlier, PSO was originally designed to optimize values in a continuous domain. However, feature selection is a binary optimization problem. Hence, to select an optimal set of features using PSO, an additional step is required to convert the continuous values into {discrete/binarized values.} For this purpose, we use a transfer function, which normalizes the optimized values to the range of $(0,1)$. We employ a standard transfer function in the domain of feature selection~\cite{chattopadhyay2021Pneumonia,ahmed2022binary}. {We generally refer to this} function as the $S-$shaped transfer {function. We present} the graphical representation of this function in Figure~\ref{fig:transfer}.
We convert the continuous values into binarized values in accordance with Equations \ref{eqn:tf} and \ref{eqn:rand}, where $rand$ is any random number in the range of $\sigma(x)$.
\begin{gather}
    \label{eqn:tf}
    \sigma (x)= \frac{1}{1+e^{-x}}\\
        \label{eqn:rand}
    S(\sigma (x)) = \begin{cases}
          1 \quad &\text{if } \, \sigma (x) \geq rand \\
          0\quad &\text{if } \, \sigma (x) < rand
     \end{cases}
\end{gather}

\subsubsection{Fitness Value}
\begin{equation}
    \label{eqn:fit}
    Fitness=\alpha \times a + (1-\alpha)\times f
\end{equation}
To evaluate the strength of candidate solutions, we define a fitness value,{ which we calculate} following Equation~\ref{eqn:fit}. In this equation, $\alpha$ is a hyperparameter, $f$ is a ratio of the number of left-out features to the total number of {features,} while $a$ is the classification accuracy. 
\subsection{Altruistic PSO}
\begin{figure}[ht!]
    \centering
    \includegraphics[width=0.8\linewidth,height=8cm,keepaspectratio]{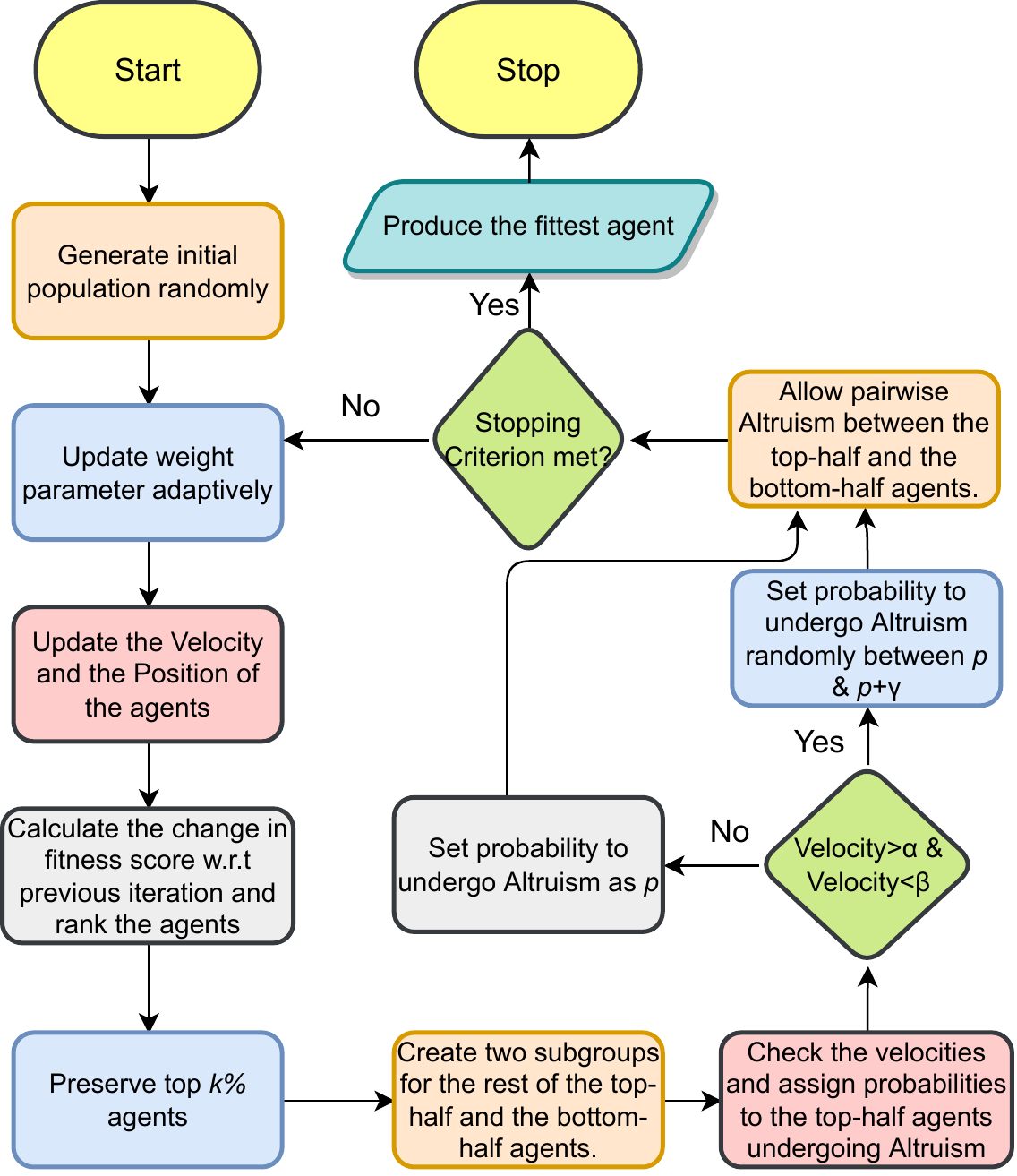}
    \caption{A flow diagram for the proposed AAPSO. The value of $\alpha$ is set as the velocity value when the probability of the feature to get selected is half ($\sigma(\alpha)=0.5$). The value of $k$ is set to $40$. Whereas the value of $\beta$ is set as the velocity value when the probability of the feature to get selected is close $0.8$ i.e., not too definite. The value of $p$ is set as $\sigma(\alpha)$ and $\gamma$ is set as $\sigma(\beta)$.}
    \label{fig:altpso}
\end{figure}

\begin{algorithm}[ht!]
      \caption{Altruism Process}
    \label{alg:altPSO}
     \begin{algorithmic}[1]
     \renewcommand{\algorithmicrequire}{\textbf{Input:}}
     \renewcommand{\algorithmicensure}{\textbf{Output:}}
     \REQUIRE $N$ agents with having a solution set $S$, each with the dimension $d$, having corresponding velocities $v$ and ranked according to the change in fitness from the previous iteration. $k$ fraction of solutions to be considered as elite.
     \ENSURE  New solution set $S$ and velocities $v$
     \\ \textit{Initialization} :
     \STATE start\_agent $\gets$ $k\times N$
     \STATE stop\_agent $\gets$ $N-k\times N$
    \FOR{$i\gets $ start\_agent to stop\_agent}
        \STATE $m$ $\gets$ $N-i$
        \FOR{$j\gets $ 1 to $d$}
            \IF{$v_{ij}>\alpha\;\&\; v_{ij}<\beta$}
                \IF{$random(0,1)< random (\sigma(\alpha),\sigma(\beta))$}
                \STATE $S_{mj}\gets$ $S_{ij}$
                \STATE $v_{mj}\gets$ $v_{ij}$
                \STATE $v_{ij}\gets$ $random(0,1)$
                \STATE $S_{ij}\gets$ Apply Equations \ref{eqn:tf} and \ref{eqn:rand} on $v_{ij}$
                \ENDIF
            \ELSE
                \IF{$random(0,1)< 0.5$}
                \STATE $S_{mj}\gets$ $S_{ij}$
                \STATE $v_{mj}\gets$ $v_{ij}$
                \STATE $v_{ij}\gets$ $random(0,1)$
                \STATE $S_{ij}\gets$ Apply Equations \ref{eqn:tf} and \ref{eqn:rand} on $v_{ij}$
                \ENDIF
            \ENDIF
        \ENDFOR
    \ENDFOR
     \end{algorithmic} 

 \end{algorithm}
For a feature selection algorithm, the strength of a solution is the ability to reduce the feature dimension. Velocities in PSO determine how this reduced feature set is generated, hence the velocities associated should be given significance when any sort of {operation is performed on them.} One major problem with {the basic} PSO is that it does not account for generational memory, or in other words, operations of the basic PSO are not affected by a change in objective function value in its previous iterations. This can be considered a limitation of the algorithm since PSO does not possess any memory other than storing the personal and global best information of the agents. This may lead the algorithm to diverge from the informative region of the search space. One of the ways to encounter this problem is to minimal change {in} the fitness value in each iteration. {Also,} from a feature selection perspective, if the probability of selection is not very definite, the selection of such a feature may not be useful for classification. For such a scenario, we can say the probability of the feature selection {process} is {neither} very definite to get accepted a feature in the reduced feature set nor to get {it} rejected as a redundant feature. {Hence, we must} give these features a chance to be re-assessed in the search {space. This might }be also beneficial as the less fit agents may be transferred to the selected feature subset.
\subsubsection{Altruism}
Altruism means showing selfless concern for the well-being of others. Both humans {and} some animals sometimes show altruistic behavior towards their family members or {friends.} This {allows} other members a chance to survive or to improve their ability to survive. To incorporate this nature into the basic PSO, we first evaluate the change in fitness scores for each agent {w.r.t.} the previous iteration and rank them accordingly. If the change is significant, in such a {scenario} we should let the agent converge to its optimal solution before using it for a possible greater advantage (in this case to explore the search space). So we preserve the elite agents (say, top $k\%$) and let them converge to the optimal solution. For the remaining agents, we allow the top-half agents to show altruistic behavior with the bottom-half pairwise (best with worst, second-best with second-worst, and so on).\\
To incorporate this idea, we consider the top-half {agents,} which undergo altruism with their partner agents. We randomly transfer the features along with their corresponding velocities to its partner. To selectively perform this task, if the value of~\ref{eqn:tf} is greater than $p$ and less than $p+\gamma$, the probability of that particular feature getting selected lies in the range $(p,p+\gamma)$. Once the whole process is over, the velocities in these dimensions, which have {undergone altruism,} are randomly reset to get optimized for a probable better solution. The overall pipeline of the AAPSO can be found in Figure~\ref{fig:altpso}. Algorithm~\ref{alg:altPSO} shows the steps of the proposed method. {We present a} dry run of the AAPSO in~\ref{ap:PSO} for the convenience of the readers. 
\section{Results and Analysis}
\label{sec:results}
In this section, we have reported the obtained results by applying our proposed method to a publicly accessible standard Pneumonia dataset. All the experiments have been performed on a machine with Nvidia Tesla T4 GPU, and the programming language used is Python 3.8. We have also performed some experiments using MATLAB programming version 9.4. We have used the PyTorch library to implement the deep learning model.
\subsection{Performance of The Deep Learner}
\begin{figure}
    \centering
    \includegraphics[width=0.7\linewidth,keepaspectratio]{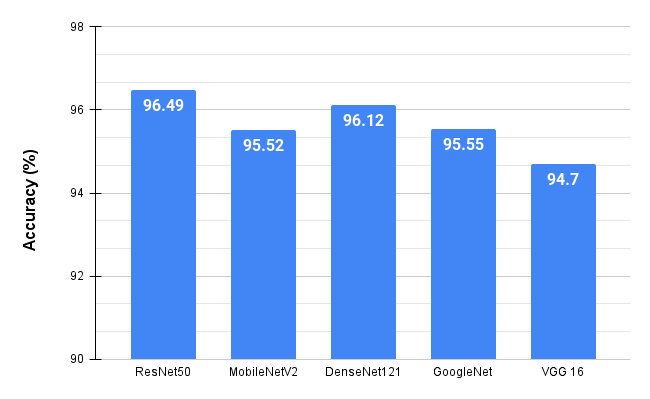}
    \caption{Classification performance of several deep learners on Fold-1 on the Pneumonia dataset}
    \label{fig:deepcompare}
\end{figure}

{Typically,} deep CNN architectures are designed to tackle some specific problems related to image classification or pattern recognition. {Therefore,} the feature extraction process is often different from one deep CNN model to another. Here, we have compared several state-of-the-art pre-trained deep CNN architectures for Pneumonia detection from CXRs. The deep models include VGG-16, one of the classic deep models which consists of 3x3 filters, and {it is 16-layers} deep. GoogleNet consists of inception modules and an auxiliary classifier. DenseNet121 consists of dense {connections,} and the MobileNetV2 adopts {depth-wise} separable convolutions. The relevant results in Figure~\ref{fig:deepcompare} clearly state the superiority of ResNet50 over other deep learners on Fold-1 of the experiment.
\subsection{Hyperparameter Tuning}
\subsubsection{Deep Learning}
Hyperparameters are one of the most {important aspects while training} a deep learner. However, it is often difficult to find the optimal state of the model to learn most efficiently. One {important} hyperparameter for a deep learner is the batch size, which defines the number of samples to {feed} to the model at {once.} Another one is the learning rate, which controls the ability of the deep model to learn. {To get the optimal values} of these, we resort to the grid search method~\cite{paul2022ensemble}. The learning rate is selected based on the results from {a} set of values $\{1e-3,1e-4,1e-5\}$ while the batch sizes considered to find the optimal hyperparameter is $\{16,32,64\}$. The final results of these experiments are provided in Table~\ref{tab:hyper}. We observe that the optimal solution is obtained for batch size equal to $32$, and learning rate equal to $1e-4$, which is decreased by a tenth factor upon completion of 5 epochs for smoother learning and to reduce overfitting. For optimization of the deep learner, the Adam optimizer is used along with the widely used cross-entropy loss.
\begin{table}[ht!]
    \centering
    \caption{Comparison of different learning rates and batch sizes on Fold-1 of the Pneumonia dataset}
    \begin{tabular}{c|c|c}
         \toprule
         Learning Rate & Batch Size & Accuracy (in \%)\\
         \midrule
         \multirow{3}{*}{1e-3} & 16 & 94.32\\
         & 32 & 96.32\\
         & 64 & 95.47\\
         \midrule
         \multirow{3}{*}{\textbf{1e-4}} & 16 & 96.12\\
         & \textbf{32} & \textbf{96.49}\\
         & 64 & 95.38\\
         \midrule
         \multirow{3}{*}{1e-5} & 16 & 93.07\\
         & 32 & 92.74\\
         & 64 & 91.62\\
         \bottomrule
    \end{tabular}
    
    \label{tab:hyper}
\end{table}
\subsubsection{Feature Selection}
{{To assess the performance of the AAPSO, we have compared it with various other optimization algorithms. Like PSO, meta-heuristic-based }feature selection algorithms require several mathematical operations to choose optimal sets of features. Consequently, the algorithms in the literature use several sets of equations, which are assisted by various parameters. These parameters have their own significance and are known to control the optimization process. Thus, it is crucial to select the right set of hyperparameters {to effectively} use these algorithms. In this work, we follow some previous methods~\cite{chattopadhyay2021Pneumonia,ahmed2022binary} in order to use the standard values of these parameters. The {parameters used} can be found in Table~\ref{tab:fspara}.}
\begin{table*}[ht!]
\caption{{Different sets of hyperparameters for various optimization algorithms used for experimentation purposes.}}
\centering
\begin{tabular}{c|cc}
\toprule
Optimization Algorithm & Parameter & Value\\
\midrule
    \multirow{3}{*}{Generic Parameters} & Population & 20\\
    & Iterations & 30\\
    & Weight for Accuracy ($\alpha$) & $\alpha$ = 0.98\\
    \midrule
    \multirow{3}{*}{GA: Genetic Algorithm} &  Gene Selection & Roulette Wheel\\
    & Crossover Probability & 0.8\\
    & Mutation Probability & 0.05\\
    \midrule
    \multirow{3}{*}{EO: Equilibrium Optimizer} & Pool Size & 4\\
    & Constants (a1,a2)& a1 = 1, a2 = 2\\
    & Generation & 0.9\\
    \midrule
    \multirow{6}{*}{MA: Mayfly Algorithm} & Attraction Constant (a1,a2) & a1=1, a2=1.5\\
    & Initial Nuptial Dance Coefficient & 0.1\\
    & Initial Random Walk Coefficient & 0.1\\
    & Gravitational Constant & 0.98\\
    & Visibility Coefficient & 2\\
    & Nupital Dance \& Random Walk updating factor ($\delta$) & 0.9\\
    \midrule
    PSO: Particle Swarm Optimization & Coefficients (r1,r2) & r1 and r2 lie in $[0,1]$\\
    \midrule
    \multirow{2}{*}{GSA: Gravitational Search Algorithm} & Initial Gravitational Constant & 6\\
    & Constant ($\epsilon$) & 0.00001\\
    \midrule
    \multirow{2}{*}{{\footnotesize GNDO: Generalized Normal Distribution Optimization}} & Lower Bound of Variables & 0\\
    & Upper Bound of Variables & 1\\
    \midrule
    \multirow{2}{*}{ASO: Atom Search Optimization} & Depth Weight & 50\\
    & Multiplier Weight & 0.2\\
    \midrule
    \multirow{2}{*}{BOA: Butterfly Optimization Algorithm} & Modular Modality & 0.01\\
    & Switch Probability & 0.8\\
    \midrule
    ALO: Ant Lion Optimizer & Antlion Selection & Roulette Wheel\\
    \midrule
    SSA: Salp Swarm Algorithm & Constants (c1,c2) & c1 and c2 lie in $[0,1]$\\
    \midrule
    \multirow{2}{*}{CSA: Crow Search Algorithm} & Awareness Probability & 0.1\\
    & Flight Length & 1.5\\
    \bottomrule
\end{tabular}
\label{tab:fspara}
\end{table*}
\subsection{Experimental Outcomes}
\begin{table}[ht!]
    \centering
    \caption{Fold wise performance of the proposed method. All metrics are reported in \%.}
    \begin{tabular}{c|cccc}
    \toprule
        Fold & Accuracy & Precision & Recall & F1  \\
        \midrule
        Fold-1 & 98.37 & 98.95 & 98.83 & 98.89\\
        Fold-2 & 98.54 & 98.95 & 99.06 & 99.00\\
        Fold-3 & 98.46 & 98.95 & 98.95 & 98.95\\
        Fold-4 & 97.86 & 97.89 & 99.17 & 98.52\\
        Fold-5 & 98.80 & 99.30 & 99.07 & 99.18\\
        \midrule
        Average & 98.41 & 98.80 & 99.02 & 98.91\\
        Standard Dev & 0.34 & 0.54 & 0.13 &0.24\\
        \bottomrule
    \end{tabular}
    \label{tab:foldwise}
\end{table}
The proposed deep feature selection based method is trained and tested using the 5-fold cross validation methodology. Table~\ref{tab:foldwise} gives fold-wise results on the dataset discussed in the preceding sub-section. One of the main reasons for not achieving cent per cent metrics lies in the fact that the dataset is quite imbalanced (see Table \ref{tab:data}). Another possible reason for {misclassification} includes high intra-class variability. For example, cases with early stages of Pneumonia which show very less prominent features in the CXRs~\cite{zhang2021viral}.
\begin{figure}[ht!]
    \centering
    \includegraphics[height=5cm]{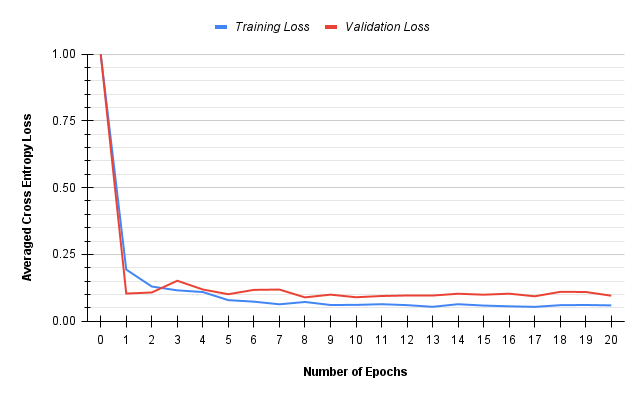}
    \caption{Learning curve w.r.t. loss for the deep learner (ResNet50). The loss values for each epoch are plotted against the progression in epoch.}
    \label{fig:loss}
\end{figure}
\begin{figure}[ht!]
    \centering
    \includegraphics[height=5cm,keepaspectratio]{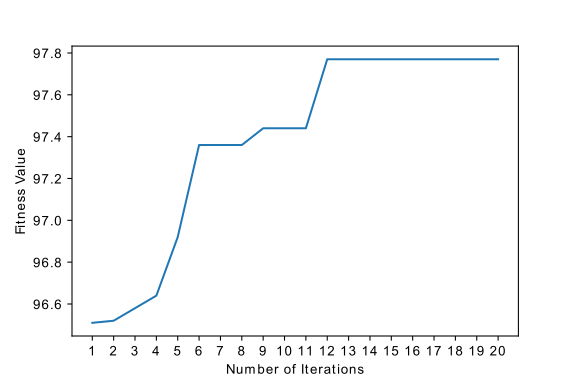}
    \caption{The learning curve for the top ranked agent in each iteration w.r.t. the progression in iteration. The X-axis refers to the progression in iteration whereas the Y-axis gives the fitness value for the top ranked agent. Note that the top ranked agent may not be the same in each iteration.}
    \label{fig:fsconv}
\end{figure}
\begin{figure}[ht!]
    \centering
    \includegraphics[height=5cm,keepaspectratio]{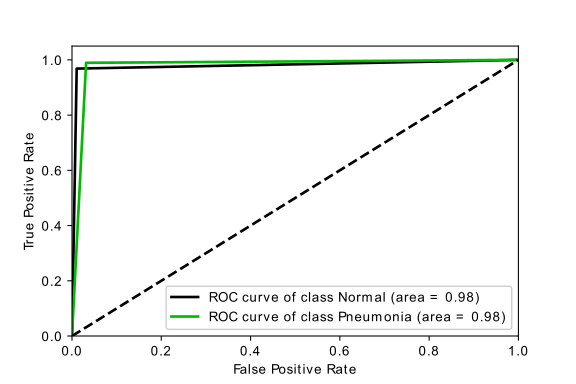}
    \caption{ROC curves for both the Pneumonia and normal classes post feature selection. The reported curves are for Fold-1 of the experiment.}
    \label{fig:roc}
\end{figure}
\begin{figure}
    \centering
    \includegraphics[height=5cm,keepaspectratio]{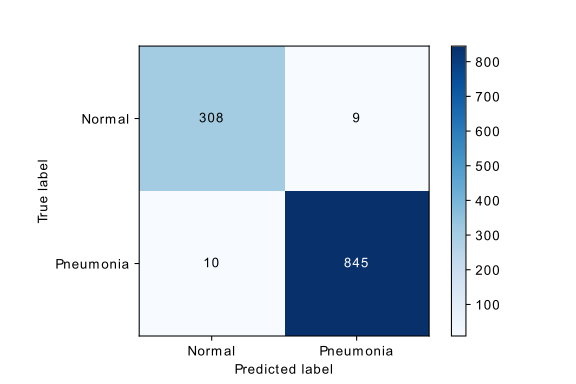}
    \caption{Confusion matrix for Fold-1 of the experiment post feature selection.}
    \label{fig:confmat}
\end{figure}
Figure~\ref{fig:loss} presents the loss w.r.t. to the epochs for the deep learner (ResNet-50). From both these figures, we observe that the CNN model does not suffer from any major overfitting. For the proposed AAPSO, we give the fitness w.r.t. to the number of iterations in Figure~\ref{fig:fsconv} which shows the stability of the algorithm as it converges over the time.
The {receiver operating characteristic} (ROC) curves in Figure \ref{fig:roc} show that the model does not behave suboptimally for any of the classes. The confusion matrix in Figure~\ref{fig:confmat} gives a quantitative measure to support this claim. 
\begin{table*}[ht!]
    \centering
    \caption{Comparison of different feature selection techniques using the 5-fold cross validation method. Acc, Feat, Avg and SD refer to accuracy (\%), number of features used to classify, average and standard deviation respectively}
    \begin{tabular}{c|cc|cc|cc|cc|cc|cc}
        \toprule 
        \multirow{2}{*}{Method} & \multicolumn{2}{c|}{Fold-1} & \multicolumn{2}{c|}{Fold-2} & \multicolumn{2}{c|}{Fold-3} & \multicolumn{2}{c|}{Fold-4} & \multicolumn{2}{c|}{Fold-5} & \multirow{2}{*}{Avg} &\multirow{2}{*}{SD} \\
        \cmidrule{2-11}
        & Acc & Feat& Acc & Feat& Acc & Feat& Acc & Feat& Acc & Feat\\
        \midrule
        ResNet50 & 96.49 & 512 &98.11 & 512 & 95.89 & 512 & 96.92 & 512 & 96.49 &512 & 96.78 & 0.83\\
        \midrule
        GA & 97.26 & 200 & 96.98 & 198 & 97.44 & 198 & 95.81 & 181 & 97.95 &169 & 97.09 & 0.80\\
        EO & 97.96 & 250 & 98.29 & 245 & 97.95 & 247 & 95.76 & 249 & 98.29 & 245 & 97.65 & 1.07\\
        MA & 96.52 & 364 & 98.46 & 370 & 97.78 & 384 & 96.07 & 348 & 97.95 & 363 & 97.36 & 1.01\\
        PSO (basic) & 97.52 &189 & 97.78 & 177 & 97.52 &193 &96.84& 194 & 98.46 & 235 & 97.62 & 0.58\\
        GSA & 98.02 &241 &97.18 &251 & 97.26 &254 &97.01 & 235 & 97.78 &274 & 97.45 & 0.43\\
        {GNDO} & 97.52 & 242 & 98.12 & 268 & 97.86 & 254 & 96.58 & 244 & 97.78 & 241 & 97.57 & 0.59\\
        {ASO} & 97.95 & 249 & 97.77 & 249 & 97.27 & 226 & 96.99 & 258 & 97.69 & 253 & 97.53 & 0.39\\
        {BOA} & 97.69 & 232 & 98.12 & 240 & 98.04 & 258 & 96.75 & 248 & 98.03 & 243 & 97.73 & 0.57\\
        {ALO} & 97.86 & 240 & 97.60 & 255 & 97.95 & 232 & 96.75 & 245 & 98.26 & 252 & 97.68 & 0.58\\
        {SSA} & 97.86 & 259 & 98.12 & 249 & 98.26 & 258 & 97.01 & 252 & 98.29 & 251 & 97.91 & 0.53\\
        {CSA} & 97.86 & 242 & 97.35 & 248 & 98.12 & 259 & 97.60 & 249 & 97.35 & 261 & 97.66 & 0.34\\
        \midrule
    Altruistic PSO & 98.03 & 173 & 98.37 & 195 & 97.95 & 191 & 96.92 & 192 & 98.12 & 189 & 97.88 & 0.56\\
        \textbf{AAPSO} & \textbf{98.37} & \textbf{163} & \textbf{98.54} & \textbf{176} & \textbf{98.46} & \textbf{183} & \textbf{97.86} & \textbf{195} & \textbf{98.80} & \textbf{183} & \textbf{98.41} & \textbf{0.34}\\
        \bottomrule
    \end{tabular}
    \label{tab:compfs}
\end{table*}
\subsection{Comparison With Other Feature Selection Algorithms}
\begin{figure}[ht!]
    \centering
    \includegraphics[height=5cm,keepaspectratio]{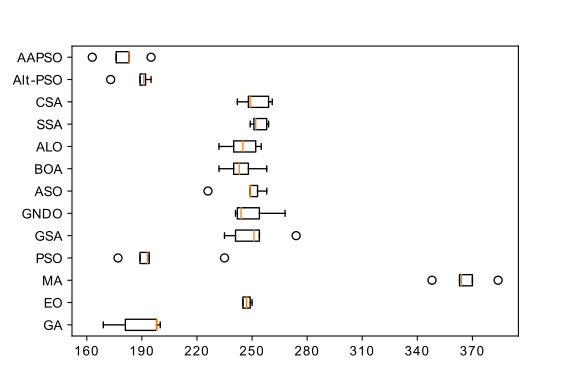}
    \caption{Box Plot analysis for the number of features. The plots were drawn using the number of features selected in 5 folds of the experiment by various feature selection algorithms.}
    \label{fig:box}
\end{figure}

{Table~\ref{tab:compfs} compares our AAPSO with 11 popular meta-heuristic based feature selection algorithms on the 5-fold cross-validation setting. The algorithms compared include GA: Genetic Algorithm, EO: Equilibrium Optimizer, MA: Mayfly Algorithm, PSO: Particle Swarm Optimization, GSA: Gravitational Search Algorithm, GNDO: Generalized Normal Distribution Optimization, ASO: Atom Search Optimization, BOA: Butterfly Optimization Algorithm, ALO: Ant Lion Optimizer, SSA: Salp Swarm Algorithm, CSA: Crow Search Algorithm. We can observe from the results that the proposed AAPSO outperforms the other methods {vis-à-vis} both in terms of classification accuracy and the number of features used. Besides, from the results shown in this table, we can also claim that the introduction of adaptive dependence on memory significantly increases the learning capability of the PSO. This increase in performance can be attributed to the fact that the adaptive parameter effectively balances proper exploration in the initial stages of the algorithm followed by exploitation in the later stages. To additionally {evaluate the stability} of the algorithm in comparison to other algorithms for different folds of data, we present the standard deviation and box-plots. From Table~\ref{tab:compfs} we observe that standard deviation values in different folds are very less which ensures the stability of the proposed AAPSO. Also,{ we show} the box-plots in Figure~\ref{fig:box} using the length of reduced feature sets obtained by the AAPSO along with other algorithms considered here for comparison. The figure shows the spread, skewness and locality of the data (in this case the number of selected features) among the group. From the data presented, we conclude that the proposed AAPSO acts optimally for each fold of {the samples} and is not biased towards any specific {set of samples} (i.e., folds). {Thus,} from this discussion, we {can} state that the AAPSO performs its intended task effectively.}
\begin{table}[ht!]
    \centering
    \caption{Results of Mann-Whitney U Test. Tests are conducted by comparing the 5-fold accuracy values produced by different feature selection algorithms.}
    \begin{tabular}{c|c}
    
    \toprule
      Method & p-value\\
      \midrule
      PSO (Basic) & 0.02293\\ 
      GA & 0.01078\\
      EO & 0.04684\\
      MA & 0.01836\\
      GSA & 0.01079\\
      GNDO & 0.01390\\
      ASO & 0.01079\\
      BOA & 0.03005\\
      ALO & 0.02327\\
      SSA & 0.03746\\
      CSA & 0.01366\\
      \bottomrule
    \end{tabular}
    \label{tab:stat_test}
\end{table}
\subsection{Statistical Analysis of AAPSO}
{We} perform a statistical significance test to determine the robust nature of the {AAPSO} algorithm when compared to other algorithms. In doing so, we consider a null hypothesis: \textit{“The proposed AAPSO provides similar results when compared to other feature selection algorithms.”} To reject this null {hypothesis}, we take the help of a very popular non-parametric statistical test, namely the Mann-Whitney U test~\cite{perez2015improving}. This test is based on the idea that two distributions namely X and Y are arranged in increasing order based on the values of X and Y. A condition is checked whether the majority of the samples in X lie above or below the majority of the samples in Y. Such a phenomenon would go against the fundamental principle of random mixing. {Hence,} the null hypothesis of random mixing would be disregarded~\cite{perez2015improving}. To gather the statistical evidence, we {consider} the classification accuracies by different feature selection algorithms for each of the five folds. If the obtained p-value is {greater} than 0.05 (5\%), we conclude the null hypothesis has enough statistical evidence to {get it accepted. Otherwise,} we reject this hypothesis. Table~\ref{tab:stat_test} presents the p-values for the aforementioned statistical test. From the tabulated results, we can reject the null hypothesis.
\subsection{Comparison With Other improved Feature Selection Algorithms}
\label{ss:impfs}
{In the preceding section, we have shown the robustness of AAPSO when compared to {various} meta-heuristic algorithms {in its basic form}. Over the past few years, researchers have also improved the algorithms to {handle} specific problems~\cite{yuan2016ttsa,kar2020fuzzy}. Therefore, {for a fair} comparison, we compare the present method with some recently proposed feature selection algorithms. The results of this comparison are recorded in Table~\ref{tab:dfsi}. {For this comparison,} we consider the following methods:
\begin{enumerate}
    \item Automata-based improved equilibrium optimizer with U-shaped transfer function~\cite{ahmed2021aieou} (AIEOU)
    \item Social ski-driver algorithm with late acceptance hill climbing~\cite{chatterjee2020late} (SSD+LAHC)
    \item Adaptive switching gray-whale optimizer~\cite{mafarja2020efficient} (ASGW)
    \item Binary Simulated Normal Distribution Optimizer~\cite{ahmed2022binary} (BSNDO)
    \item Embedded chaotic whale survival algorithm for filter-wrapper feature selection~\cite{guha2020embedded} (ECWSA)
\end{enumerate}
From Table~\ref{tab:dfsi} it is clear that AAPSO outperforms the improvised feature selection algorithms considered here.}
\begin{table}[ht!]
    \centering
    \caption{Comparison of AAPSO with improvised feature selection algorithms on Fold-1 of the experiment}
    \begin{tabular}{c|cc}
    \toprule
        Algorithm & Accuracy (\%) & Features \\
        \midrule
       AIEOU & 98.02 & 202\\
       SSD+LAHC	& 98.21 & 198\\
       ASGW & 97.98 & 187\\
       BSNDO & 98.32 & 189\\
       ECWSA & 97.65 & 243\\
       \midrule
       Alt-PSO & 98.03 & 173\\
       AAPSO & 98.37 & 163\\
       \bottomrule
    \end{tabular}
    
    \label{tab:dfsi}
\end{table}
\subsection{Comparison of Various Deep Learning Models with AAPSO-based Feature Selection}
{To {ensure} the proper {usage} of the feature selection algorithm, we should ensure that the proposed method is not biased towards the deep learning model considered here (i.e., ResNet50). One effective way to check this is how the proposed method performs when other standard deep learning models are used. For this purpose, we use four state-of-the-art deep learning models other than ResNet50. As described in the previous sections, following the same {methodology}, we extract the features and apply our feature selection algorithm.{ We provide the relevant} results of this experiment in Table~\ref{tab:deeppso}.}
\begin{table}[ht!]
    \centering
    \caption{Ablation study considering various deep learning models along with ResNet50 for Fold-1 of the experiments. Here Acc (\%) and Feat represent the classification accuracy and the number of selected features respectively}
    \begin{tabular}{c|c|cc|cc|cc}
    \toprule
    \multirow{2}{*}{Deep Model} &\multirow{2}{*}{Acc} & \multicolumn{2}{c|}{PSO} & \multicolumn{2}{c|}{Altruistic PSO} & \multicolumn{2}{c}{AAPSO}\\
    \cmidrule{3-8}
    & & Acc & Feat& Acc & Feat& Acc & Feat\\
    \midrule
    VGG16 & 94.70 & 97.01 & 186 & 97.18 & 242 & 97.61 & 227\\
    GoogleNet & 95.55 & 96.67 & 220 & 97.09 & 211 & 97.35 & 221\\
    DenseNet121 & 96.12 & 97.44 & 227 & 97.42 & 220 & 97.53 & 231\\
    MobileNetV2 & 95.52 & 95.73 & 224 & 96.24 & 217 & 96.50 & 194\\
    \midrule
    ResNet50 & 96.49 & 97.52 &189 & 98.03 & 173 & 98.37 & 163\\
    \bottomrule
    \end{tabular}
    \label{tab:deeppso}
\end{table}
{From Table~\ref{tab:deeppso}, it is clear that the proposed AAPSO performs well with all the deep models. We can observe a significant increase in the classification accuracy, while the performance of the ResNet50 model is considerably better in terms of classification {accuracy. In addition,} we also see post AAPSO feature-selection the accuracy is better than applying AAPSO feature selection {with other deep learners.}{ We may attribute this} success to the ability of ResNet50 to generate a better feature representation than other models for the classification problem under consideration. This is {because} while training a deep model with a neural network {classifier,} the classifier intends to form linearly separable features for better hyperplane formation~\cite{an2015can}.}
\subsection{Comparison with State-of-the-Art Methods}
\begin{table}[ht!]
    \centering
    \caption{Comparison with state-of-the-art methods. Values of the performance metrics are shown in terms of \%}
    \begin{adjustbox}{max width=\linewidth}
    \begin{tabular}{c|cccc}
    \toprule
        Work Ref. & Accuracy & Precision & Recall & F1  \\
        \midrule
        Linag \& Zheng~\cite{liang2020transfer} & 90.50 & 89.10 & 96.70& 92.70\\
        Sharma et al.~\cite{sharma2020feature}& 90.68 & - & - & -\\
        Stephen et al.~\cite{stephen2019efficient} & 93.73 & - & - & -\\
        Ibrahim et al.~\cite{ibrahim2021Pneumonia} & 94.43 & - & 98.19 & -\\
        Saraiva et al.~\cite{saraiva2019classification} & 95.30 & 98.86 & 94.77 & 96.77\\
        Rajaraman et al.~\cite{rajaraman2018visualization} & 96.20 & 97.70 & 96.20 & 97.00\\
        Dey et al.~\cite{dey2021customized} & 97.94& 95.02& 97.55& 96.27\\
        Mahmud et al.~\cite{mahmud2020covxnet} & 98.10 & 98.00& 98.50& 98.30\\
        Chattopadhyay et al.~\cite{chattopadhyay2021Pneumonia} & 98.36 & 98.98 & 98.79 & 98.88\\
        \midrule
        \textbf{Ours} & \textbf{98.41} & \textbf{98.80} & \textbf{99.02} & \textbf{98.91}\\
        \bottomrule
    \end{tabular}
    \end{adjustbox}
    \label{tab:sota}
\end{table}
While evaluating any approach, it is always important to compare {the proposed approach} with the {recent approaches} found in the literature. Table~\ref{tab:sota} compares the proposed approach with the existing approaches in terms of various performance metrics. From the tabulated {results,} we observe that our method outperforms the state-of-the-art approaches. This increase can attribute to the proposed enrichment in PSO to select relevant subsets of features from a given feature set generated by the ResNet50 model.
\subsection{Further Analysis}
\begin{figure*}[ht!]
    \centering
    \includegraphics[width=0.73\linewidth,height=4cm]{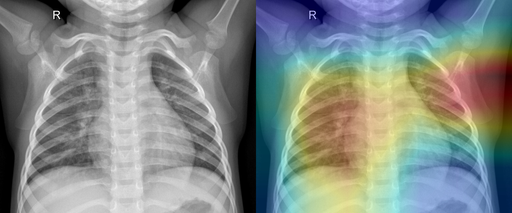}
    \includegraphics[width=0.73\linewidth,height=4cm]{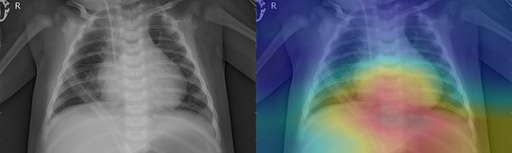}
    \caption{GradCAM analysis on testing data. The upper figure shows the GradCAM analysis for a normal image with classification probability of 0.9875. The bottom image shows GradCAM analysis for a confirmed Pneumonia case with classification probability 0.9965. The GradCAMs are generated using the gradient maps in the final convolutional layer of the deep learner.}
    \label{fig:gradcam}
\end{figure*}
The role of a feature selection algorithm is to primarily select a relevant set of features from the entire set of features and then further classify based on the selected features. However, it becomes an important task to extract relevant features. Figure~\ref{fig:gradcam} presents Gradient-weighted class activation maps (GradCAM) for two {cases,} one being the normal case and the other being a confirmed Pneumonia case. The normal case as expected does not have a strong activation in any region but has an overall activation in the entire chest area. This is very important since this implies the model relies on the full CXR image for its final prediction. In contrast, we see a strong activation for the confirmed Pneumonia case near the lower region close to the heart. One possible reason which is very common in the case of Pneumonia is that due to {gravity,} water deposits in the lungs settle in the bottom region and thus form bacterial colonies in that region~\cite{aly2008randomized}. From a detection perspective, the strong activation of the model in such regions is of great {significance,} which the model does effectively as seen in the figure. However, we also see some areas lie outside the region of interest, which in this case is the chest region. {These} areas have {slightly medium to less} activation in both these figures. This is when feature selection helps to increase the performance of the model. A feature selection algorithm aims to discard such redundant features, which may degrade the classification performance. Thus, from an overall perspective, it becomes an important task to identify the proper features and design suitable techniques to overcome the shortcomings of the feature extraction processes.
\section{Performance of AAPSO on other real-life datasets}\label{sec:real}
{Feature {selection-based} algorithms are often {devised} to solve real-life problems more efficiently and effectively. To ensure this quality of a feature selection algorithm, we apply AAPSO to some real-life datasets frequently {used} in the literature. We consider datasets from the UCI repository, microarray-based gene expression datasets and a COVID-19 prediction dataset. For comparison, we choose some existing improvised feature selection algorithms as described in Section~\ref{ss:impfs}. The reason behind such a choice is that the respective algorithms have already shown their effectiveness when compared to classical algorithms in the respective papers. {Also,} this section compares the results with AAPSO {only,} as the preceding section establishes the superiority of AAPSO over {Altruistic PSO}. In this section, we {furthermore} mention some strengths and weaknesses of AAPSO and suggest some potential solutions to overcome the said weaknesses. This is required to cope up with the growing demand for efficient artificial intelligence (AI)-powered systems.}
\subsection{AAPSO on UCI Datasets}
{As stated earlier, to test the effectiveness of AAPSO, {we conduct} experiments on several standard datasets obtained from UCI repository~\footnote{https://archive.ics.uci.edu/ml/index.php}. The details of {the datasets} can be found in Table~\ref{tab:ucides}. We tabulate the comparative results in Table~\ref{tab:uci}.}
\begin{table}[ht!]
    \centering
    \caption{{Details of the datasets chosen from the UCI machine learning repository.}}
    \begin{tabular}{c|ccc|c}
    \toprule
    \multirow{2}{*}{Dataset} & \multicolumn{3}{c|}{Description} & \multirow{2}{*}{Domain}\\
    \cmidrule{2-4}
       & Attributes & Samples & Classes & \\
       \midrule
Breastcancer&	9&	699&	2&	Biology\\
BreastEW&	30&	569&	2&	Biology\\
Exactly2&	13&	1000&	2&	Biology\\
HeartEW&	13&	270&	2&	Biology\\
IonosphereEW&	34&	351&	2&	Electromagnetic\\
KrvskpEW&	36&	3196&	2&	Game\\
Lymphography&	18&	148&	4&	Biology\\
SonarEW&	60&	208&	2&	Biology\\
SpectEW&	22&	267&	2&	Biology\\
WineEW&	13&	178&	3&	Chemistry\\
\bottomrule
    \end{tabular}
    
    \label{tab:ucides}
\end{table}

\begin{table}[ht!]
    \centering
    \caption{{Results on standard UCI datasets. Acc (\%) and Feat refer to the classification accuracy and the number of features selected respectively. Note that the bold values in AAPSO refer to the best metric for the dataset amongst competitors}}
    \label{tab:uci}
   \begin{tabular}{c|cc|cc|cc|cc|cc|cc}
   \toprule
     \multirow{2}{*}{Dataset} & \multicolumn{2}{c|}{AAPSO} & \multicolumn{2}{c|}{AIEOU}& \multicolumn{2}{c|}{BSNDO}& \multicolumn{2}{c|}{SSD+LAHC}& \multicolumn{2}{c|}{ECWSA}& \multicolumn{2}{c}{ASGW}\\
    \cmidrule{2-13}
& Acc & Feat & Acc & Feat & Acc & Feat & Acc & Feat & Acc & Feat & Acc & Feat\\
    \midrule
Breastcancer&	\textbf{100}&	\textbf{3}&	100&	8&	100&	4&	98.93&	3&	95.21&	7&	98.50&	5	\\
BreastEW&	\textbf{100}&	11&	98.25&	3&	98.25&	4&	98.25&	9&	97.38&	15&	100&	16	\\
Exactly2&	\textbf{81.15}&	\textbf{6}&	80.50&	8&	80.50&	8&	79&	8&	78.90&	9&	77.70&	8	\\
HeartEW&	85.18&	5&	90.74&	4&	90.74&	4&	91.67&	5&	85.63&	9&	83.1&	6	\\
IonosphereEW&	90.15&	9&	95.74&	11&	95.74&	16&	96.43&	12&	86.79&	10&	97.2&	17	\\
KrvskpEW&	\textbf{100}&	25&	99.53&	6&	98.44&	22&	97.81&	20&	93.53&	16&	97.1&	25	\\
Lymphography&	\textbf{96.67}&	\textbf{6}&	96.67&	6&	96.67&	5&	96.67&	7&	87.02&	10&	88.40&	11	\\
SonarEW&	95.24&	26&	95.24&	6&	95.24&	27&	97.62&	24&	76.84&	23&	94.80&	36	\\
SpectEW&	\textbf{98.15}&	8&	98.15&	14&	96.22&	6&	95.15&	9&	79.84&	7&	87&	10	\\
WineEW&	\textbf{100}&	4&	100&	3&	100&	3&	100&	3&	98.02&	7&	100&	6	\\
   \bottomrule
    \end{tabular}
\end{table}
{From Table~\ref{tab:uci} we observe that AAPSO performs the best in terms of classification accuracy for 7 out of 10 datasets. While the number of features reduced may not be individually best, results are comparable, thereby ensuring the effectiveness of AAPSO for real-world problems.\\
AIEOU is an improvised version of EO, with {automata-based learning}. The method uses Adaptive $\beta$ Hill Climbing $(A\beta HC)$ to find a better {equilibrium} pool. The parameters of EO {were} made adaptive using a 3-action automata. BSNDO is an improvised version of GNDO coupled with simulated annealing for better local search. {Similarly,} SSD+LAHC couples Social-Ski Diver (SSD) with Late acceptance hill-climbing (LAHC) local search. {However,} these methods do not use {data-driven strategies} but focus on a better adaptive balance between exploration and exploitation. ECWSA uses a filter-wrapper-based approach for Whale Optimization Algorithm (WOA) the selection was guided by chaos. ASGW uses an adaptive agent switching strategy for a hybrid Grey Wolf Optimizer (GWO) and WOA. {These} methods focus on intelligent agent switching strategies, which the authors balance between exploration and exploitation. However, there was no explicit focus on controlling the agent's exploration and exploitation strategies.\\
While we see the effectiveness of AAPSO on UCI datasets, {we} should {note} that UCI datasets have comparatively fewer features than we have considered for the deep feature selection framework. Thus, we also conclude that AAPSO performs well even in low dimension settings.}
\subsection{AAPSO on High Dimensional Data}
{While we have observed the effectiveness of AAPSO on deep feature selection and standard UCI {datasets,} this subsection examines the ability of AAPSO to handle very high dimensional datasets. For this purpose, {we consider} three microarray datasets namely DLBCL, Prostate and SRBCT. All these datasets consist of gene expressions. These datasets are used {for} cancer prediction. Handling microarray data is very challenging because the search space is very huge. We {observe} the results of this experiment in Table~\ref{tab:microarray}.}
\begin{table}[ht!]
    \centering
    \caption{{Results on very high dimensional microarray datasets. The accuracy score (\%) is presented under the method, whereas the number of selected features is written in brackets.}}
    \begin{tabular}{c|c|ccccccc}
    \toprule
    Dataset &Features& AAPSO	& AIEOU	& BSNDO	& SSD+LAHC	& ECWSA	& ASGW\\	
    \midrule
    DLBCL & 7070 & \textbf{100(114)} & 100(182) & 100(162) & 96(98) & 100(154) & 94(214)\\
    Prostate & 12,533 & 100(212) & 95(34) & 100(201) & 100(284) & 96(178) & \textbf{100(98)}\\
    SRBCT & 2308 & \textbf{100(214}) & 100(236) & 100(298) & 94(68) & 100(265) & 100(231)\\
    \bottomrule
    \end{tabular}
    
    \label{tab:microarray}
\end{table}
{The results in Table~\ref{tab:microarray} clearly {showcase} the reliability of AAPSO on real-life cancer prediction from given sequences of genes. Thus, AAPSO is also useful for high dimensional data.}
\subsection{AAPSO on COVID-19 Prediction}
{Undoubtedly, COVID-19 had a traumatizing impact on humanity since its emergence in late 2019. Early diagnosis of COVID-19 is always encouraged by medical professionals and governmental and non-governmental agencies to curb the spread of the virus. The early symptoms include loss of smell, shortness of breath, and fever {amongst} many others. Since these symptoms are so common for any other viral diseases, there should be some specialized way to testing.\\
These symptoms are generally treated as features and are fed to a suitable classifier for detecting the presence of the said virus. One such dataset is publicly available~\footnote{\href{https://github.com/Atharva-Peshkar/Covid-19-Patient-Health-Analytics}{https://github.com/Atharva-Peshkar/Covid-19-Patient-Health-Analytics}} for development of AI-powered COVID-19 detection tools. This dataset has 74 such features {and} 1085 instances.}
\begin{table}[ht!]
\caption{{Results on COVID-19 dataset. The accuracy score (\%) is presented under the method, whereas the number of selected features is written in brackets.}}
    \centering
    \begin{tabular}{c|cccccc}
    \toprule
         Features& AAPSO& AIEOU	& BSNDO	& SSD+LAHC	& ECWSA	& ASGW\\
         \midrule
         74 & \textbf{99.08(29)} & 96.31(52) & 98.61(26) & 97.69(23) &94.31(50) & 97.69(40)\\
         \bottomrule
    \end{tabular}
    
    \label{tab:covid}
\end{table}
{Table~\ref{tab:covid} shows a comparative analysis for COVID-19 detection. Based on these results we can mention that AAPSO works well even for COVID-19 prediction.}
\subsection{Strengths, Weaknesses, and Future Extension of AAPSO}
{To get an unbiased view of the {AAPSO,} in this subsection we summarize the strengths and weaknesses of AAPSO. In addition, we also provide some suitable suggestions to tackle the problems faced by AAPSO.}
\subsubsection{Strengths}
{
\begin{itemize}
    \item AAPSO undergoes selective altruism by preserving the top ranked agents, thereby allowing them to converge to the optimal solution.
    \item The altruism helps to explore certain subsets of features, i.e., the lower ranked agents, in order to give them a chance to improve their fitness value.
    \item The adaptive behavior helps to set an adaptive balance between exploration and exploitation, thus providing a way for the algorithm to look into the search space in an effective manner.
\end{itemize}}
\subsubsection{Weaknesses}
{
\begin{itemize}
    \item AAPSO requires a bit more computation than the basic PSO, thus requiring additional computational resources.
    \item The number of agents should be preferably higher $(>10)$ to have top-ranked agents to be preserved and have a suitable number of agents to undergo altruism. This leads to a greater training time.
    \item The altruism is static in nature, or in other words, altruism happens to a fixed number of agents regardless of their performance. This may not be useful for certain cases.  
\end{itemize}
}
\subsubsection{Future Extension}
{\begin{itemize}
    \item The number of agents undergoing altruism may be selected dynamically based on some statistical measures.
    \item We may consider the past information of an agent to keep a track if the agent is trapped in local optima or not. This may increase the memory use, but can be effective in certain scenarios.
    \item We may consider having guided initialization strategies to help explore the search space in a more computationally efficient manner, thus reducing the overall time of execution.
\end{itemize}}
\section{Conclusion}
\label{sec:conclusion}
While the world heals from the effects of the devastating COVID-19 pandemic, still Pneumonia concerns us with its fatality rate and other consequences. In this work, we have proposed an AI-based technique for automatic Pneumonia detection from CXRs. 
For this task, we have considered a pre-trained base deep CNN learner namely ResNet50 and fine-tuned it on a standard Pneumonia dataset. We have extracted features from the second last layer and {employed} the proposed AAPSO for feature selection and classified {the CXRs} based on the selected features. Extensive experiments and thorough analysis establish the robustness of the proposed method when compared to some state-of-the-art methods.\\
In future, we aim to incorporate some dynamic characteristics to further increase the performance of PSO by efficiently balancing its exploration and exploitation capabilities. We also plan to incorporate other types of data-driven altruistic strategies into PSO or other optimization algorithms. Further work can also focus on designing very lightweight deep learners which will be {useful for deploying} in resource-constrained environments. {We can also plan to incorporate noise reduction techniques~\cite{bi2019temporal} to enhance the quality of the raw image and make it more informative.}
\section*{CRediT Authorship Contribution Statement}
\textbf{Rishav Pramanik:} Conceptualization, Methodology, Software, Investigation, Validation, Writing—Original Draft, Writing—Review \& Editing
\textbf{Sourodip Sarkar:} Software, Investigation, Writing—Original Draft 
\textbf{Ram Sarkar:} Conceptualization, Methodology, Investigation, Visualization, Resources, Writing—Review \& Editing, Supervision, Project administration
\section*{Acknowledgements}
The authors thank the Center for Microprocessor Applications for Training Education and Research (CMATER) research laboratory of the Computer Science and Engineering Department, Jadavpur University, Kolkata, India for providing infrastructural support to this work.
\bibliographystyle{elsarticle-num} 
\bibliography{manuscript}

\appendix
\section{Dry run of AAPSO}
\label{ap:PSO}
Consider a dry run of AAPSO with $3$ agents in its $15^{th}$ iteration, where the maximum number of iterations is $30$ and the number of dimensions of each agent is $5$. Let us assume that we preserve the top $33\%$ of the solutions during altruism. Let $V$ be the velocities and $X$ be the solution set, where 1 refers to a selected feature and 0 refers to a discarded feature.
\[
V=
  \begin{bmatrix}
    -10.25 & 0.36 & 12.02 & 1.25 & -5.65 \\
    15.65 & -3.65 & 5.12 & 0.12 & 2.65 \\
    5.32 & -3.12 & 2.62 & -1.25 & 5.85\\
  \end{bmatrix}\]
\[
X= \begin{bmatrix}
  0 & 0 & 1 & 1 & 0\\
  1 & 0 & 1 & 1 & 1\\
  1 & 0 & 1 & 0 & 1\\
\end{bmatrix}
\]
The steps go like:
\begin{itemize}
    \item \textbf{Step-1: }Update the adaption parameter, $w=1-e^{-0.5}=0.393$
    \item \textbf{Step-2: }Update the velocities according to Equation~\ref{eqn:newPSO}. \emph{This is an assumption that all agents achieve their personal best in this iteration only. While in real-life scenarios, this may or may not be the case.}\\
    The updated velocities for the $i^{th}$ look like: $V_i=0.393*V_i+0.69*(V_i-X_i)+0.42*(V_1-X_i)$ and the feature set is updated. Note the values $0.69 \& 0.42$ are randomly generated:
    \[
V=
  \begin{bmatrix}
    -15.41 & 0.54 & 16.95 & 0.77 & -8.49 \\
    11.53 & -3.80 & 9.48 & -0.45 & -0.61 \\
    0.35 & -3.23 & 2.62 & -0.83 & 2.86\\
  \end{bmatrix}
  \]
\[
X= \begin{bmatrix}
  0 & 1 & 1 & 1 & 0\\
  1 & 0 & 1 & 0 & 0\\
  0 & 0 & 1 & 0 & 1\\
\end{bmatrix}
\]  
    \item \textbf{Step 3: }Calculate the change in fitness w.r.t. the previous iteration and rank them accordingly. Let us \underline{consider} the ranking is the order $X_2,X_1,X_3$. Therefore, as per the steps described in the main sections, $X_2$ is preserved while $X_1\;\&\;X_3$ undergo altruism.
    \item \textbf{Step 4:} Perform altruism between top-half non-elite solutions and bottom-half non-elite solutions. The updated velocity and feature set are:
    \[
V=
  \begin{bmatrix}
    \underline{0.63} & 0.54 & 16.95 & \underline{0.31} & \underline{0.89} \\
    11.53 & -3.80 & 9.48 & -0.45 & -0.61 \\
    \textbf{-15.41} & -3.23 & 2.62 & \textbf{0.77} & \textbf{-8.49}\\
  \end{bmatrix}\]
\[
X= \begin{bmatrix}
  \underline{0} & 1 & 1 & \underline{0} & \underline{1}\\
  1 & 0 & 1 & 0 & 0\\
  \textbf{0} & 0 & 1 & \textbf{1} & \textbf{0}\\
\end{bmatrix}
\]
The bold values indicate inheritance from altruistic partner, while the underlined values signify a random number for resetting.
\item \textbf{Step 5: }Evaluate the feature set and go for the next iteration
\end{itemize}
\section*{Vitae}
\begin{wrapfigure}{l}{25mm} 
    \includegraphics[width=1in,height=1.25in,clip,keepaspectratio]{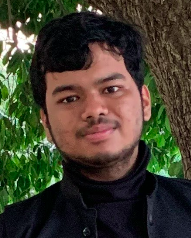}
  \end{wrapfigure}\par
  \textbf{Rishav Pramanik} is presently an undergraduate student of the Department of Computer Science and Engineering at Jadavpur University, Kolkata, India. He is a researcher at the undergraduate level and keeps himself up-to-date by reading several journals and conference proceedings on a regular basis. His research interests include Deep Learning, Image and Video Processing and Nature Inspired Optimizers.\\~\\~\\~\\\par
\begin{wrapfigure}{l}{25mm} 
    \includegraphics[width=1in,height=1.25in,clip,keepaspectratio]{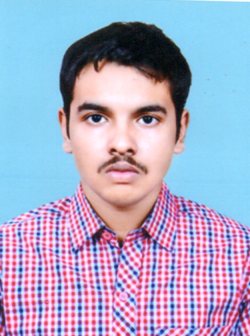}
\end{wrapfigure}\par
\textbf{Sourodip Sarkar} is presently an undergraduate student of Electronics and Communication Engineering department at Heritage Institute of Technology, Kolkata, India. He is passionate about problem-solving by computer softwares. His research interests comprise Image Processing, Pattern Recognition, Machine Learning and Analytics.\\~\\~\\~\\~\\\par

\begin{wrapfigure}{l}{25mm} 
    \includegraphics[width=1in,height=1.25in,clip,keepaspectratio]{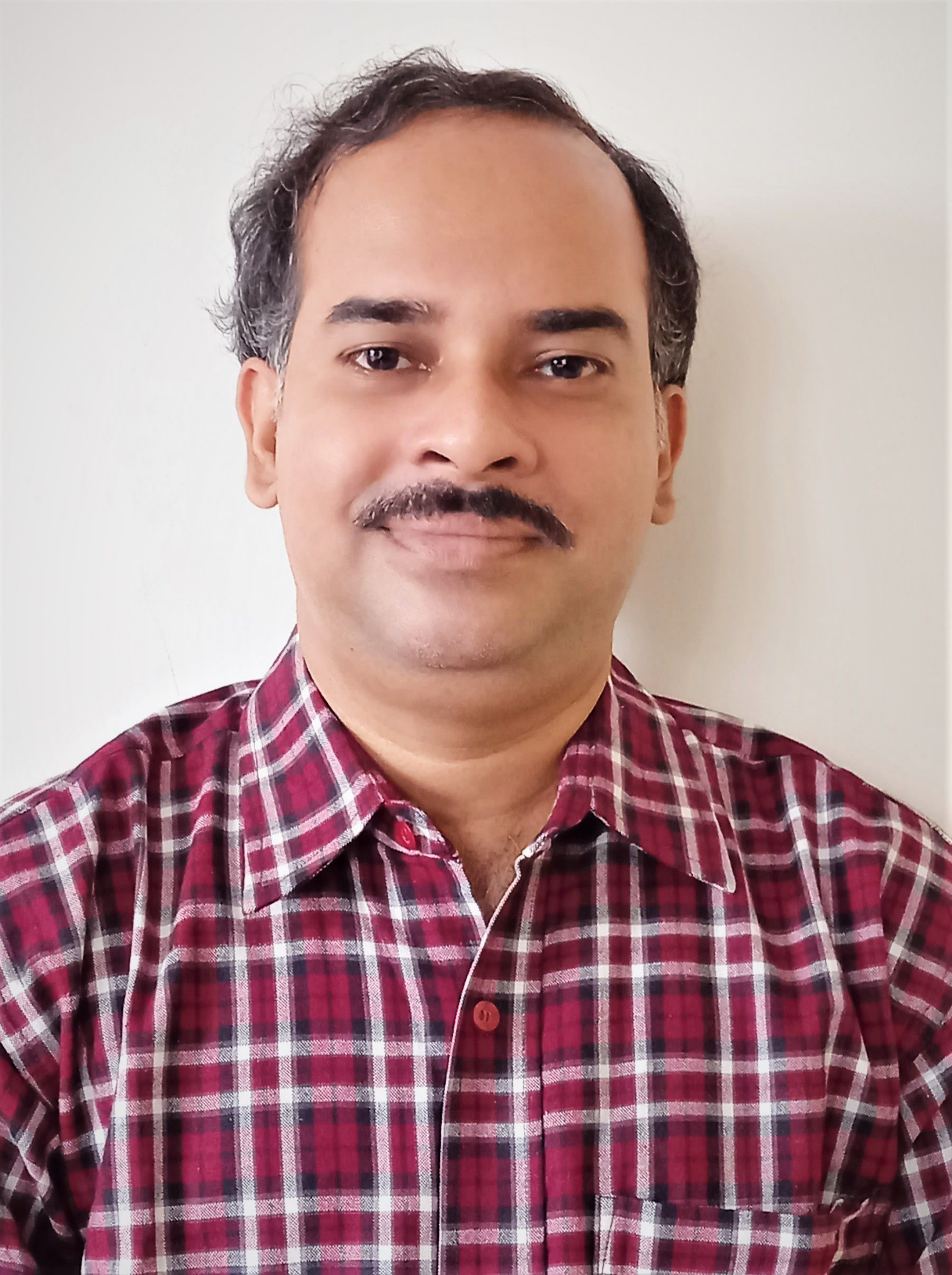}
  \end{wrapfigure}\par
  \textbf{Ram Sarkar} received his B. Tech degree in Computer Science and Engineering from University of Calcutta, India in 2003. He received his M.E. degree in Computer Science and Engineering and PhD (Engineering) degree from Jadavpur University, Kolkata, India in 2005 and 2012 respectively. He joined the department of Computer Science and Engineering of Jadavpur University as an Assistant Professor in 2008, where he is now working as a full Professor. He received the Fulbright-Nehru Fellowship (USIEF) for post-doctoral research in University of Maryland, College Park, USA in 2014-15. He has published more than 300 research papers in various Journals and Conference Proceedings. His research areas include Image and Video Processing, Optimization Algorithms and Computer Vision. He is a senior member of IEEE and a member of ACM.\par
\end{document}